\documentclass{article}
\usepackage[utf8]{inputenc} % allow utf-8 input
\usepackage[T1]{fontenc}    % use 8-bit T1 fonts
\usepackage{hyperref}       % hyperlinks
\usepackage{url}            % simple URL typesetting
\usepackage{booktabs}       % professional-quality tables
\usepackage{amsfonts}       % blackboard math symbols
\usepackage{nicefrac}       % compact symbols for 1/2, etc.
\usepackage{microtype}      % microtypography
\usepackage{graphicx}

\graphicspath{{./img/}}

\usepackage{combelow}
\usepackage{paralist}
\usepackage{booktabs}

\title{Protecting Retail Investors from Order Book Spoofing using a GRU-based Detection Model}

\author{%
  Jean-Noël Tuccella\thanks{\texttt{jean-nol.tuccella19@imperial.ac.uk}} \\
  Department of Computing\\
  Imperial College London\\
  London SW7 2AZ, UK
  \and
  Philip Nadler\thanks{\texttt{p.nadler@imperial.ac.uk}} \\
  Data Science Institute\\
  Imperial College London\\
  London SW7 2AZ, UK
  \and
  Ovidiu \cb Serban \thanks{ \texttt{o.serban@imperial.ac.uk}}\\
  Data Science Institute\\
  Imperial College London\\
  London SW7 2AZ, UK 
}

\begin{document}

\maketitle

\begin{abstract}
Market manipulation is tackled through regulation in traditional markets because of its detrimental effect on market efficiency and many participating financial actors. The recent increase of private retail investors due to new low-fee platforms and new asset classes such as decentralised digital currencies has increased the number of vulnerable actors due to lack of institutional sophistication and strong regulation. This paper proposes a method to detect illicit activity and inform investors on spoofing attempts, a well-known market manipulation technique. Our framework is based on a highly extendable Gated Recurrent Unit (GRU) model and allows the inclusion of market variables that can explain spoofing and potentially other illicit activities. The model is tested on granular order book data, in one of the most unregulated markets prone to spoofing with a large number of non-institutional traders. The results show that the model is performing well in an early detection context, allowing the identification of spoofing attempts soon enough to allow investors to react. This is the first step to a fully comprehensive model that will protect investors in various unregulated trading environments and regulators to identify illicit activity.
\end{abstract}

%PN: We need to ensure to incorporate their topics in our paper to make it sexy

    % Systemic bias and its impact on financial outcomes on different customer segments 
    % Metrics of fairness 
    % Auditing the disparate impact of credit decisioning and lending
    % Theories of equal treatment and impact
    % Understanding and controlling machine learning biases
    % Enforcing fairness at training time
    % The relationship between fairness theory and fair lending regulation
    % Market fairness and metrics of market fairness-
    % AI/ML for fair market regulations 
    % Fairness in human-in-the-loop systems
    % Decisions under uncertainty and fairness
    % Explainability techniques for investigating bias and fairness
    % Impacts of decision complexity on fairness outcomes

%PN:
%TODO
% Figures need better/more accurate description. A reader should understand the meaning of a figure by just looking at it and the underlying description, without having read the maintext

%Model Training Details

%Some more references

%broken references

\section{Introduction }
\label{sec:intro}

Price manipulations occurs when someone is trying to trick other investors on the value of an asset listed on an exchange platform. This can take various forms: wash-trading, pump-and-dumping, spoofing, etc. Most of these techniques are illegal in all regulated exchanges, as an example these practices were made  illegal in the US by the Dodd-Frank Act of July 2010 \cite{sar2017dodd}. However, most of the cryptocurrency exchanges are not regulated and are often accused to turn a blind eye or even participate in these frauds. One recent example has been highlighted by a study  in\cite{griffin2020}, accusing Bitfinex (one of the largest cryptocurrency exchanges) and its related entities Tether (USDT-stablecoin issuing company) to be responsible for a massive price manipulation in 2017.

In an unregulated ecosystem without any central legal authority, investors should not blindly trust any entity and should always verify the environment allowing them to invest in crypto-assets. This paper fits in this logic by first informing on price manipulations occurring on crypto-exchanges and then, by paving the way to efficient protection for investors.

This project's main objective is the detection of price manipulations in several cryptocurrency exchanges to protect investors from fraud. In particular, this paper focuses primarily on spoofing, but the techniques presented can be easily adapted to detect other illicit activity. Spoofing is the act of placing a large order on the market with no actual plan to execute it. This makes other investors believe that the market has reached a consensus on the manipulated price, thus shifting the market up or down.

Our goal is to first identify spoofing events and allow us to label the time series in two categories: manipulated or not. Second, a Gated Recurrent Unit (GRU)-based architecture (\cite{chung2014empirical}) will be tested on the classification of the time series. Fast detection of illicit activity allow for fast reactions and potential protection from losses due to fraud, our work will also focus on the earliness of detection as a performance indicator. The model will have to classify time series before the instant that could lead to a loss for retail investors.

\subsection{Spoofing description}
\label{sec:spoof_def}
Illicit activity, due to its nature, is generally hard to define so rather than attempting a formal definition we will be giving a description of a very common scenario. A fraudulent investor could act as follow:
\begin{itemize}
    \item He will first place one or multiple large orders on one side of the order book, as close as possible to the current bid-ask to maximise the impact on the market \cite{lee2013microstructure}. This would be the fake order that he is not willing to execute.
    \item If the spoofing succeeds, the orders will trigger a move on the real market price.
    \item Then the trader will cancel the previously placed orders, usually just before someone starts executing it.
    \item He will finally place orders on the other side of the order book, thus taking advantage of the market shift.
\end{itemize}

A more detailed description of the phenomenon can be found in Figure 3, Section D of Leangarun et al. paper \cite{leangarun2016stock}.

In order to be able to detect this type of activity, the data processing algorithm will have to have access to:
\begin{inparaenum}
    \item the volume of the orders,
    \item  the distance of the price of orders to the current market bid-ask and
    \item the volatility of prices.
\end{inparaenum}

%PN: This should be called market data, since we are not making an on-chain/off-chain market data split.
\subsection{Market data}

The market data currently used by this project can be divided in two levels: 
\begin{inparaenum}
    \item Level 1 data is the most basic: it contains Opening/Highest/Lowest/Closing prices and the volume traded in a given period.
    \item Level 2 data contains more order book information, which is usually composed of all bid and ask orders that have not been yet matched by the exchange.
\end{inparaenum}

For this research project, we are using primarily level 2 data for several pairs of cryptocurrency tokens such as: BTC/USD, DAI/USD, BTC/DAI from several exchanges including Bitfinex and Kraken.
The data has been gathered over several months (November 2019-May 2020) on a milliseconds frequency, which totals to 337GB of raw uncompressed data.

\section{Related work}

\subsection{Unsupervised anomaly detection}
The problem of price manipulation spotting can be seen as the detection of anomalies in trading signals. In an unsupervised framework, the time series are not labelled. The model will first learn a representation of the data. If future data points are too far from the representation, the model will consider it as an anomaly.

In \cite{gunnemann2014robust}, the authors try to detect anomalies in the rating of products and services on websites such as Amazon, TripAdvisor, or Yelp. They create a model that learns the base behaviour of users by considering it as a latent multivariate auto-regressive process (Robust Latent Autoregression or RLA). The observation set being already polluted with anomalies, it is not possible to directly infer the base behaviour of users. The authors, therefore, consider that the observed time series are the mix of the base and abnormal behaviour. Then, if the deviation between newly seen observation and the base behaviour of users is too large, an anomaly is detected.

A major advantage of this method is that it does not require labelled data. However, clustering data implies the loss of temporal information which is critical to our problem. Furthermore, spoofing is a well documented type of fraud. As a consequence, labelling our time series based on the definition of spoofing is possible to do without major effort. 

\subsection{Supervised Deep Learning}
The anomaly detection problem can also be seen as a classification task where time series needs to be separated into two categories based on whatever the price is manipulated or not.

In \cite{leangarun2016stock}, the authors focus on a way to detect spoofing and pump-and-dumping (buy stock to push its price up before dumping all the stock). They use a simple feed-forward neural network on labelled level 1 data from NASDAQ companies (Amazon, Intel, and Microsoft). Unfortunately, the model was not able to detect spoofing. The reason is that level 1 data does not contain enough information, thus preventing the neural network to learn the manipulation mechanism.

In time series analysis, Recurrent Neural Networks (RNN) are widely used because they can grasp temporal dependencies in sequences of data. State of the art recurrent neural network such as GRU (Gated Recurrent Units) \cite{cho2014learning} can therefore show strong performances in time series classification problems. Elsayed et al. \cite{elsayed2018deep} shows that GRU mixed with Fully Convolutional Neural Networks can yield good performances on the classification of different types of time series. This architecture is still outperforming state-of-the-art models in many univariate time series classification problems.

\section{Data Processing and Labelling}

\subsection{Spoofing Conditions}
\label{sec:conditions}
In our current setup, the detection model is developed as a supervised learning problem. The time series are labelled and the model will try to predict the category of the time series, either manipulated or not. We follow the same approach as Leangarun et al. \cite{leangarun2016stock} to label the L2 data using thresholds. 

Based on the previously established definition of spoofing, the following conditions are defined on the state of the order book to detect price manipulation:

\textit{\underline{Condition 1 :}}

As seen in Section \ref{sec:spoof_def}, the volume of the cancelled order needs to be large enough. We can compare the volume of each cancelled order to the aggregated volume of the order book for the 25 top price levels.

\begin{itemize}
    \item $V^{bid/ask}_{cancelled}(t) \ \ \ $ : Volume of the cancelled order
    \item $V^{cumul (bid/ask)}_{25}$ : Cumulative volume of the 25 top price levels  (bid or ask side).
\end{itemize}

\begin{equation}
    condition_1 = V^{bid/ask}_{cancelled} > threshold_1*V^{cumul (bid/ask)}_{25}
\end{equation}

\textit{\underline{Condition 2:}}

The price level of the cancelled order needs to be close enough to the current bid-ask. The distance signal needs therefore to be under a certain threshold:

\begin{equation}
    condition_2 = \delta_{cancelled}^{bid/ask} > threshold_2
\end{equation}

\textit{\underline{Condition 3:}}
The successful spoofing manipulation leads to an increase in price volatility. As a consequence, the third condition is based on the $\Delta_\sigma$ signal: 

\begin{equation}
    condition_3 = \Delta_\sigma > threshold\_3
\end{equation}

\par 
A spoofing order must therefore meet all these conditions: 
\begin{equation}
    condition_{1}\ \&\ condition_{2}\ \&\ condition_{3} = True
\end{equation}

\subsection{Detection}
\label{sec:spoof_detection}
The following time series is an example of an alleged spoofing attempt detected thanks to the previously established conditions on the 9th of March 2020 on the ETH/USD market of the Kraken exchange. 

\begin{figure}[htb]
  \centering
    \includegraphics[width=1\textwidth]{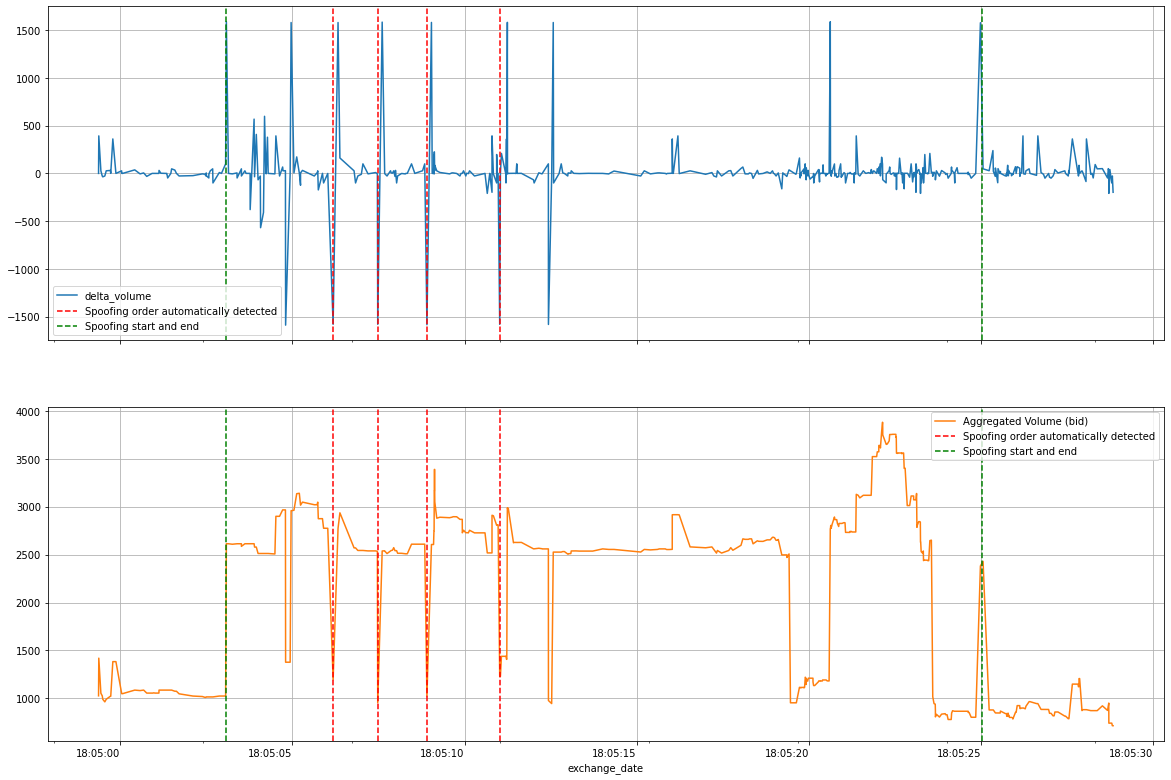}
    \caption{Volume of orders and Aggregated volume of the order book (bid side). Red vertical markers represent detected spoofing attempts by using the thresholds described in Section \ref{sec:conditions}. Some orders were missed and they were manually annotated, with the full spoofing event found between the green lines.}
    \label{fig:delta_volume}
\end{figure}

In Figure \ref{fig:delta_volume}, four consecutive suspicious orders are flagged, all containing a very large volume of ETH (1500 ETH worth approx. 150 000 USD at the time). These large orders are added then cancelled several times on the bid side. The aggregated volume of the bid side of the order book is therefore experiencing sudden changes. The regularity in the patterns seems to indicate that this is the work of one wealthy individual or entity (ETH whale) using a trading bot with very high frequency trading capability. In this case the added orders were cancelled immediately after placing them.

It also appears that the manipulation is not restricted to the 4 detected orders. The algorithm (described in Section \ref{sec:conditions}) misses suspicious orders at the beginning and at the end of the time series. Therefore, the start and the end of the price manipulation have been here manually annotated, with the full spoofing event visible between the green lines on Figure \ref{fig:delta_volume}.

\begin{figure}[htb]
  \centering
    \includegraphics[width=0.85\textwidth]{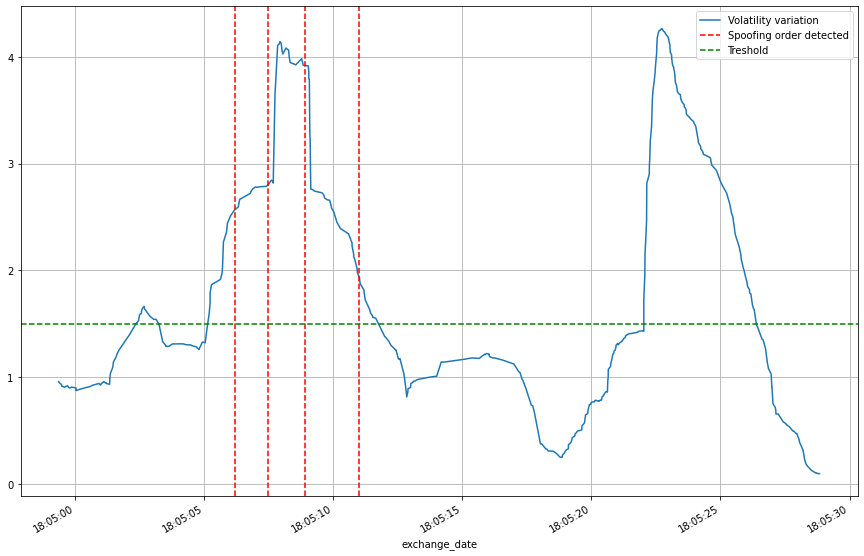}
    \caption{Price volatility variation during the spoofing attempt described in Figure \ref{fig:delta_volume}}
    \label{fig:vol_var}
\end{figure}

The price level shown on the figure is very close to the current bid, at about 1\% price move. Therefore, these orders are likely to have an impact on the current ETH price. This can be verified by looking at Figure \ref{fig:vol_var} showing the volatility variation  of the prices.

As expected, the volatility variation is substantially increasing and stays above the threshold for several seconds. This can be seen also by looking at the mid price for the same period shown in Figure \ref{fig:mid_price}: 

\begin{figure}[htb]
  \centering
    \includegraphics[width=0.85\textwidth]{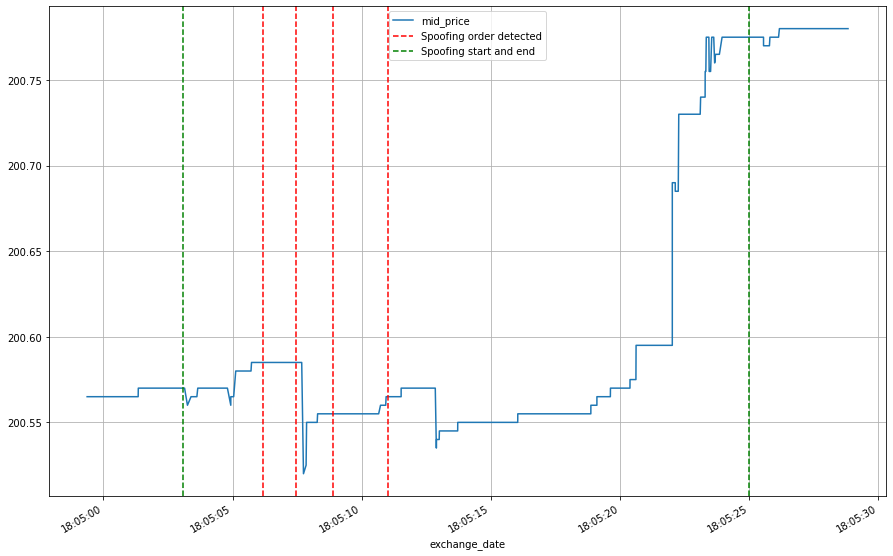}
    \caption{Mid price variation during the spoofing attempt described in Figure \ref{fig:delta_volume}}
    \label{fig:mid_price}
\end{figure}

The large orders are placed on the bid side, attempting to artificially drive up the price. The plots above show that the price is increasing by 20 cents (USD) in the suspicious 30 seconds periods.

The end of the manipulation, where prices move the most, is unfortunately not detected automatically. This is because no cancellation of orders is sent to the exchange. As we can see on Delta Volume shown on Figure \ref{fig:delta_volume}, the two last orders are sent to the exchange without being cancelled. However, changes in the aggregated volume of the order book can still be seen. This can be potentially explained by a wash-trading attempt, which is not studied by the current paper. This makes other investors believe that a real trade has been successfully executed, whereas in reality nothing happened.

\section{Early Detection Model}

Retail investors are often victims of price manipulation because they cannot react in time when transactions are happening. The spoofer has already cancelled the fake large volume order and took advantage of the price move. Our proposed model is dealing with this problem by building an early detection model that will spot price manipulations soon enough to allow investors to react. The features considered for this paper are also prioritised to allow early detection, rather than high accuracy after the spoofing event.

For all our experiments we use features extracted from high-level L2 data. We are also training the model to detect the price manipulation before the event ends to achieve the early detection effect. During annotation, we considered that the price manipulation is finished when all the labelling conditions described in Section \ref{sec:conditions} are met.\newline
%PN: We can also move this to marketdata section 1.2
The frequency of updates received by the exchanges is very high (1 every 0.03 sec for Bitfinex on BTC/USD pair) which leads to a very large amount of data to analyse (300 million price level updates for Bitfinex BTC/USD). We achieved the order book computation by updating a dictionary representing the volume of each price level for the bid and ask side. These updates are done following very simple rules (remove, add, update) and lead to a new state of the order book.To increase computational efficiency, the depth of the computed order book is limited to 25 price levels.

\begin{figure}[htb]
  \centering
    \includegraphics[width=0.8\textwidth]{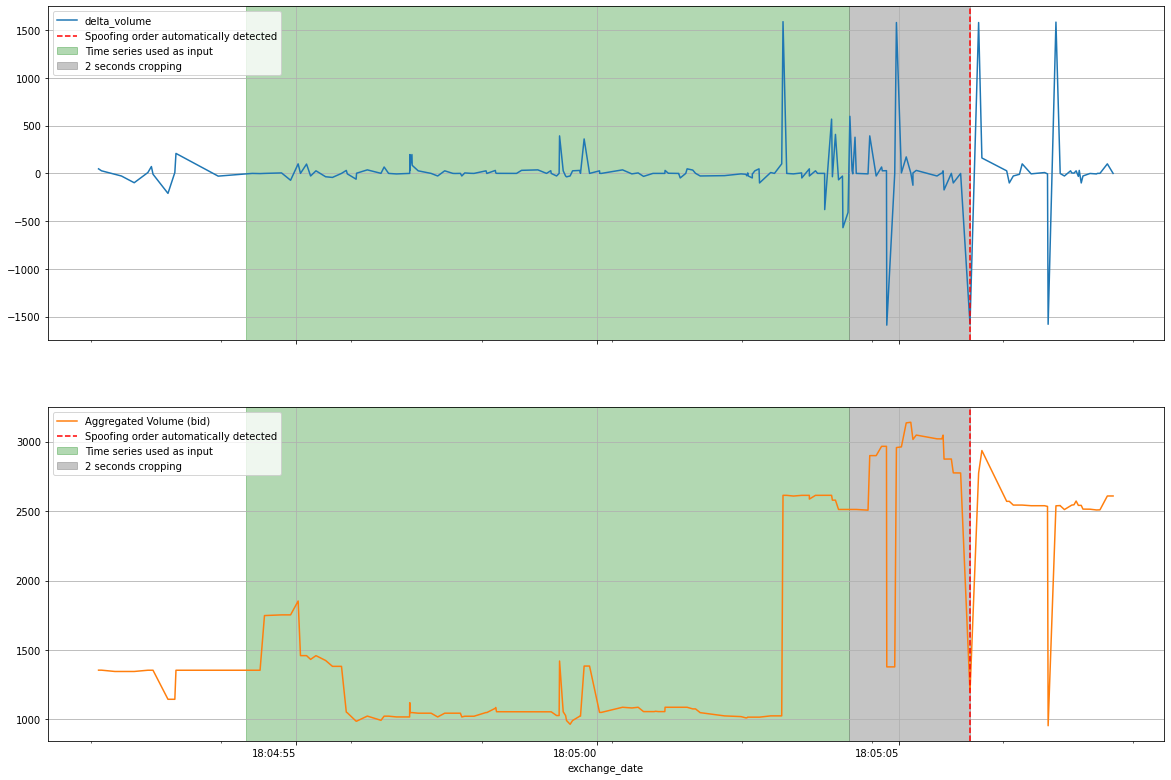}
    \caption{The early detection protocol used by our experiments}
    \label{fig:crop}
\end{figure}

The time series used by this paper consist of 200 updates of the state of the order book, which is divided in two categories: manipulated or not. The manipulated time series are ending with the order that met all labelling conditions of section \ref{sec:conditions} (i.e. the large volume cancelled order). The 2 last seconds of these time series (including the large volume cancelled order) are then removed. Figure \ref{fig:crop} illustrates the process: 1. an order is flagged at time $T_0$ (red dot on the plot) by the labelling conditions; 2. all data points between $T_0 - 12$ seconds and $T_0 - 2$ seconds (green area) are used to train the classifier model as a manipulated sample. 

By using this protocol, on a successful prediction, the model is capable of predicting manipulation 2 seconds before it ends. That could give enough time to other investors to react and avoid executing orders on the manipulated prices. 

\subsection{Classification Model Architecture}

The price manipulation mechanisms are reflected in time series by complex temporal patterns, but Recurrent Neural Networks (RNN) are powerful models that allow grasping such temporal dependencies. The model used in this paper will be based on Gated Recurrent Units (GRU) \cite{chung2014empirical} as they share the same performance of classic Long Short-Term Memory Networks (LSTM) \cite{hochreiter1997long} but are less complex, thus resulting in lower training time. This is a key feature important for our detection framework as it needs to be able to be trainable and deploy-able as quick as possible \cite{chung2014empirical}.

\begin{figure}[htb]
  \centering
    \includegraphics[width=1\textwidth]{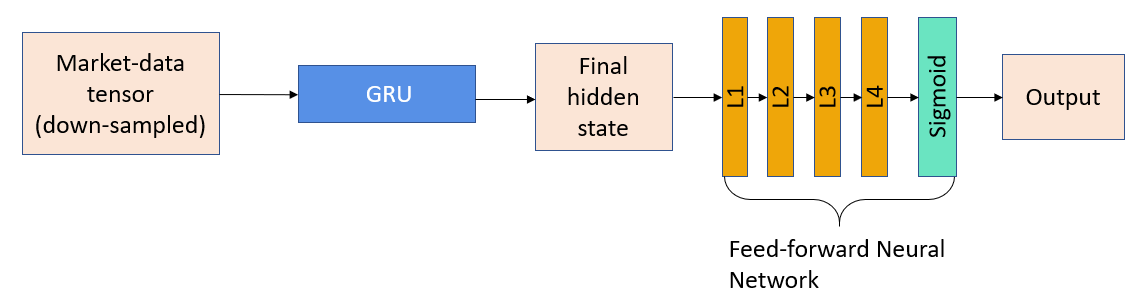}
    \caption{Classification model architecture}
    \label{fig:classification-model}
\end{figure}

 The GRU output is used to train a feed-forward neural network. The final layer of the neural network will be a scalar value corresponding to the probability of the time series being manipulated or not. Figure \ref{fig:classification-model} shows a graphical representation of our model\footnote{The full model is available at: \url{BlindedForReview}.}.
The model has been trained using the Adam optimiser due to its computational efficiency on large datasets \cite{kingma2014adam}. A learning rate decay was also implemented to improve the stability of the Binary Cross-entropy Loss. The hyperparameters (i.e. number of GRU layers, learning rate, dropout, number of training epochs) were tuned using a random search strategy.

The chosen network architecture is flexible and allows the integration of other signals and features, such as prices of other assets, other exchange specific information (traded volumes, average prices among other currencies, etc.). To prove this flexibility claim, we experimented with different architectures for additional features, such as cryptocurrency tokens being sent to an exchange at their network address, which can be potential strategy of arbitrageurs. The details of this architecture are in the Appendix, Figure \ref{fig:classification-model-bypass}. Ultimately, for our experiments we selected the architecture presented in Figure \ref{fig:classification-model} as a good balance between the accuracy, pre-processing complexity and training time. 

\subsection{Experiments}

The dataset used for all our experiments consists in 4 annotated samples following the conditions described in Section \ref{sec:conditions}. Each sample describes a different style of price manipulation, varying in term of distribution of the features, length of the event and sampling frequency. The features used for these experiments are standard L2 signals, such as:
\begin{inparaenum}
    \item current bid-ask,
    \item delta volume,
    \item price of the current price level update,
    \item price volatility,
    \item cumulative volume (bids), and
    \item cumulative volume (asks).
\end{inparaenum}
Each sample is split into training, validation, and test set (65\%, 15\%, and 20\%). The validation set is used to fine tune the GRU hyperparameters. Table \ref{tab:sample_description} shows each sample characteristics, including the sampling frequency.

\begin{table}[htb]
\centering
  \begin{tabular}{lrrrr}
        \toprule
           & sample 1 & sample 2 & sample 3 & sample 4 \\
          \midrule
          Training (down-sampled) & 44/88 & 62/124 & 176/352 & 107/214 \\
          Validation & 6/1118 & 11/1115 & 66/1098 & 15/1118 \\
          Test & 29/1493 & 15/1488 & 122/1465 & 19/1491 \\
          Overall frequency & 1.1\% & 1.2\% & 4.8\% & 1.2\% \\
          \bottomrule
          \addlinespace
  \end{tabular}
\caption{Sample characteristics, including the frequencies of manipulated time series}
\label{tab:sample_description}
\end{table}

Given that all the samples have different data distributions, for all the experiments, the model will be trained and validated on one sample, and then tested on everything else (e.g. train the model on sample 1 and test on samples 1, 2, 3, 4). This should also show the model's potential to generalise across samples. To compensate for any training bias, the models are trained 5 times and the average accuracy is reported. We also use a weighted accuracy to compensate for class imbalance. Table \ref{tab:results_experiment} shows these results. The proposed model could be evaluated against a rule-based system (i.e. Section \ref{sec:conditions}) or any other system that detects spoofing events after the fact, but it would be unfair as these systems are not capable or early detection. 

\begin{table}[htb]
\centering
  \begin{tabular}{l | rrrrr}
        \toprule
         & \multicolumn{5}{c}{\textbf{Test sample}} \\
          \textbf{Train sample} & sample 1 & sample 2 & sample 3 & sample 4 & Average \\
          \midrule
          sample 1 & 0.86 & 0.58 & 0.60 & 0.80 & 0.71 \\
          sample 2 & 0.85& 0.66& 0.58 & 0.76 & 0.71 \\
          sample 3& 0.55& 0.55 & 0.62& 0.58& 0.58\\
          sample 4& 0.83& 0.53& 0.64& 0.83 & 0.71\\
          \midrule
          Average & 0.77 &0.58 &0.61 &0.74 & 0.68 \\
          \bottomrule
          \addlinespace
  \end{tabular}
\caption{The classification Accuracy  on four separate samples}
\label{tab:results_experiment}
\end{table}

It can be noted that, when the train and test data distributions are the same, the test accuracy is higher, except sample 3. All models, except the one trained on sample 3, reach the highest accuracy when tested on sample 1. This may be because the order execution pattern from other samples generalised well for sample 1 as well.

This experiment shows that the performance of the model depends on two things : 
\begin{inparaenum}
    \item The training set: model trained with sample 2 has the best overall accuracy whereas the one trained with sample 3 has a very poor predictive power.
    \item The test set: all models tested on sample 2 and 3 reach poor accuracy. 
\end{inparaenum}

To smooth these results and hopefully yield better accuracy, the next experiment will consider the concatenation of all 4 training sets and validation sets. As performed in the previous experiment, the model will then be tested on each separate sample. The results of the experiment are shown in Table \ref{tab:detailed_results}.

\begin{table}[htb]
\centering
  \begin{tabular}{lrrrrr}
         \toprule
          & sample 1 & sample 2 & sample 3 & sample 4 & Average \\
          \midrule
          FP &0.14 & 0.68 & 0.32  & 0.24 & 0.35 \\ 
          FN &  0.10 & 0.10 & 0.32  &  0.13 & 0.16 \\
          Accuracy &0.88 & 0.61  & 0.68 & 0.81 & 0.75 \\
          \bottomrule
          \addlinespace
  \end{tabular}
\caption{The False Positive Rate (FP), False Negative Rate (FN) and Accuracy  on the four separate samples}
\label{tab:detailed_results}
\end{table}

The average performance on all samples is better than the first experiment. The model manages, almost all the time, to reach a better accuracy per sample. The model reaches good performances for sample 1 and sample 4 (almost 90\% for sample 1).
However, for sample 2 the false positive rate is very high compared with other samples. The same thing can be seen in the false-negative rate for sample 3. 
This leads to an average accuracy of \textbf{75\%} which remains a good result.

All these experiments show that it is possible to detect with good confidence a price manipulation two seconds before it ends. Unfortunately, there are still scenarios where the model still struggles to predict price manipulations. When testing on sample 2, the False Positive Rate is still too large (68\%) and may lead to many false alarm triggers. In scenario 3, the False Negative Rate is also quite high, leaving investors exposed to manipulations.

\section{Conclusion}
The aim of this project is to inform on market manipulations occurring on cryptocurrency exchanges and to build a proof-of-concept of efficient protection for retail investors. To achieve these goals we propose a pre-processing technique and a classification model that can be used for early detection. 
Market data extracted from various cryptocurrency exchanges are used to build the state of the order book of a specific market at any given time. Following that, predictive trading signals have been computed. For our experiments we selected 4 samples which describe well the spoofing phenomenon. The results show a promising start for early detection of illicit activity on cryptocurrency exchanges, with a good accuracy and a model capable of performing the classification in real-time, if needed.

% The goal was to use a supervised learning framework to teach the model how to recognize market manipulation patterns in trading signals. A model based on high-level data has been able to perform early detection of market manipulations with good accuracy in some situations. It can detect market manipulations in advance to protect investors. Some suggestions to improve the model can also been made: training on more data and increasing the complexity of the model could have a positive impact. This would allow the model to reach a better accuracy and consistency in results, thus resulting in even better protection from market manipulations.

\bibliographystyle{unsrt}
\bibliography{references}

\newpage
\begin{appendix}
\section{Appendix}

\begin{figure}[htb]
    \centering
    \includegraphics[width=0.8\textwidth]{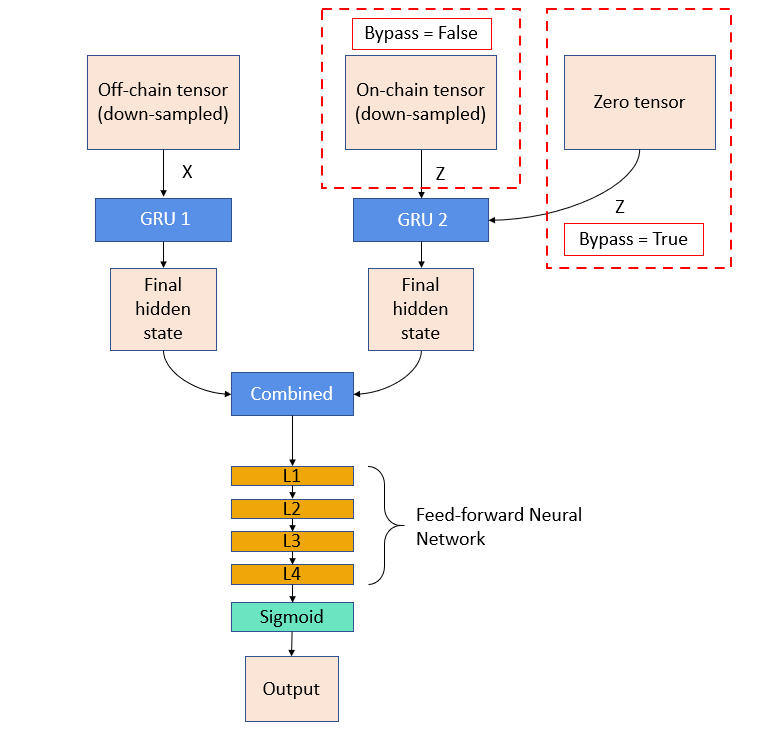}
    \caption{Example for a model extension using additional data from other sources such as on-chain network transactions, activating and deactivating the bypass component allows for comparison of forecasting accuracies when including additional signals.}
    \label{fig:classification-model-bypass}
\end{figure}

\end{appendix}

\end{document}